# Citing and referencing habits in Medicine and Social Sciences journals in 2019

Erika Alves dos Santos – School of Communication and Arts (ECA), Department of Information & Culture (CBD), University of São Paulo (USP), Brazil / Fundação Jorge Duprat Figueiredo de Segurança e Medicina do Trabalho (Fundacentro), Brazil / Digital Humanities Advanced Research Centre (DHARC), Department of Classical Philology and Italian Studies, University of Bologna, Italy

Silvio Peroni – Digital Humanities Advanced Research Centre (DHARC), Department of Classical Philology and Italian Studies, University of Bologna, Italy / Research Centre for Open Scholarly Metadata, Department of Classical Philology and Italian Studies, University of Bologna, Italy

Marcos Luiz Mucheroni – School of Communication and Arts (ECA), Department of Information & Culture (CBD), University of São Paulo (USP), Brazil

## Abstract

**Purpose** – This article explores citing and referencing systems in Social Sciences and Medicine articles from different theoretical and practical perspectives, considering bibliographic references as a facet of descriptive representation.

**Design/methodology/approach** – The analysis of citing and referencing elements (i.e. bibliographic references, mentions, quotations, and respective in-text reference pointers) identified citing and referencing habits within disciplines under consideration and errors occurring over the long term as stated by previous studies now expanded. Future expected trends of information retrieval from bibliographic metadata was gathered by approaching these referencing elements from the FRBR Entities concepts.

**Findings** – Reference styles do not fully accomplish with their role of guiding authors and publishers on providing concise and well-structured bibliographic metadata within bibliographic references. Trends on representative description revision suggest a predicted distancing on the ways information is approached by bibliographic references and bibliographic catalogs adopting FRBR concepts, including the description levels adopted by each of them under the perspective of the FRBR Entities concept.

**Research limitations** – This study was based on a subset of Medicine and Social Sciences articles published in 2019 and, therefore, it may not be taken as a final and broad coverage. Future studies expanding these approaches to other disciplines and chronological periods are encouraged.

**Originality** – By approaching citing and referencing issues as descriptive representation's facets, findings on this study may encourage further studies that will support Information Science and Computer Science on providing tools to become bibliographic metadata description simpler, better structured and more efficient facing the revision of descriptive representation actually in progress.

**Paper type** Research paper

# Introduction

Providing bibliographic information to make documents searchable and retrievable through several access paths using structured metadata corresponds to one of the purposes of *descriptive representation*. Descriptive representation is a process in which we identify data related to the editorial production of documents, such as their authors, titles, publishers, years of publication, number of pages, sources, and other bibliographic elements (Galvão, 1998; Lancaster, 2004). Although descriptive representation is still considered the physical description in manual catalogs (Maimone *et al.*, 2011), it is also a way to materialize and make evident information's meaning, context, and relationships (Lancaster, 2004) between documents. One of the products of this type of information representation is the bibliographic reference (Galvão, 1998), considered by Baptista (2007) as the elaboration of records containing the descriptive representation.

Bibliographic references are a facet of descriptive representation. They correspond to a specific kind of description that acts like a reference that points the reader to an original source of information cited by an author in the text body of a work. Each bibliographic reference contains the textual representation of a minimal set of descriptive bibliographic metadata which enables the identification of a publication, a speech, a piece of information, or anything else that may be citable, to locate and retrieve it (ABNT, 2018; Cunha and Cavalcanti, 2008; ISO, 2010).

This article aims to study bibliographic references and the other contextual entities – namely mentions, quotations, and in-text reference pointers[1], summarized in Figure 1 and Figure 2 – from several perspectives.

Baptista (2007) argues that bibliographic references are now being written by different professionals, according to their interests – scientists, artists, enterprises, negotiators, publishers, libraries, archives, museums, etc. – coming from multiple locations. Bibliographic references are one of the tools that establish links between scholarly works and enable the creation of citation networks. Writing bibliographic references correctly demands previous background and some Information Science skills. It requires that both reference and citation styles guidelines are presented in comprehensive language, with a wide scope of the bibliographic universe, especially considering Baptista's statement. Information Science should have a primary role in this activity. However, since researchers in all the disciplines usually deal directly with this aspect without involving experts, one can introduce mistakes in the bibliographic references that prevent the clearly and unambiguously identification of the represented (i.e. cited) works.

Citation error is an old issue from the XVII Century (Sweetland, 1989), and started before the publication of the first reference style manuals – the Hart's Rules for Compositors and Readers (1893) and the University of Chicago Manual of Style (1906) (O'Connor, 1977). Despite the publication of those style manuals, "citation errors continued to appear, as did an increasing number of complaints about them" (Sweetland, 1989, p. 293).

Although Sweetland's study was published decades ago, his arguments remain updated and most of his conclusions are remarkably similar to the findings of our study suggesting that neither the publication of reference styles nor the technological advances were effective in solving those bibliographic issues.

---

[1] For the sake of clarity, the definitions of all the bibliographic terms mentioned above and the others used in this study can be found in Santos *et al.* (2020a).

**Table 1.** The main citing and referencing errors pointed by Sweetland in his study (1989)

| Core citing and referencing errors | Possible reasons and justifications for the errors |
|---|---|
| Lack of standardization in citation formats | • While based on particular standard formats, multiple versions of references styles seem to have made little improvement in citation accuracy.<br>• Errors are made in the first place because the "complexity" or lack of standardization. Given the variety of formats of citation and the lack of any real agreement among journals or authors, the chance of misunderstandings is high.<br>• Lack of standardization, noting the variants in authors names; spelling tendency to invert vowels and number pairs; miscited page numbers, incorrect and misleading journal titles and wildly misspelled author's names, incorrect and incomplete citations; lacking or incorrect work's titles, use of the same abbreviations to refer to two different journals; cite only one reference when two or more were listed under a single number in the cited article, miscopy numbers.<br>• Complaints about lack of uniformity about authors or librarians. |
| Diffusion of responsibility in the publishing process | • The role of citations is not taken very seriously by the scientific community.<br>• Errors are not discovered and corrected before publication. |
| General lack of training in the norms of citation | • General human inabilities to reproduce long strings of information correctly.<br>• Tendency to invert vowels and number pairs.<br>• Misunderstanding of foreign languages. |
| Failure to examine the document cited | • Citers have not actually seen the original work.<br>• Dishonesty: fake experimental data or references to inexistent papers. |

Starting from the premise that the problems referred by Sweetland still stand, the purpose of this article is to reflect on how recent journal articles published worldwide in all the subject categories of two main academic disciplines (i.e. Medicine and Social Science) organize their citation apparatus. We focus on how bibliographic references are defined in published articles and the way they are denoted within the article text through the related in-text reference pointers to support the author's argumentation.

**Figure 1.** Visual representation of the main bibliographic elements involved in citing and referencing author-date system – i.e. when in-text reference pointers referring to mentions and quotations include author's surnames and year of publication of the cited works described in a particular bibliographic reference.

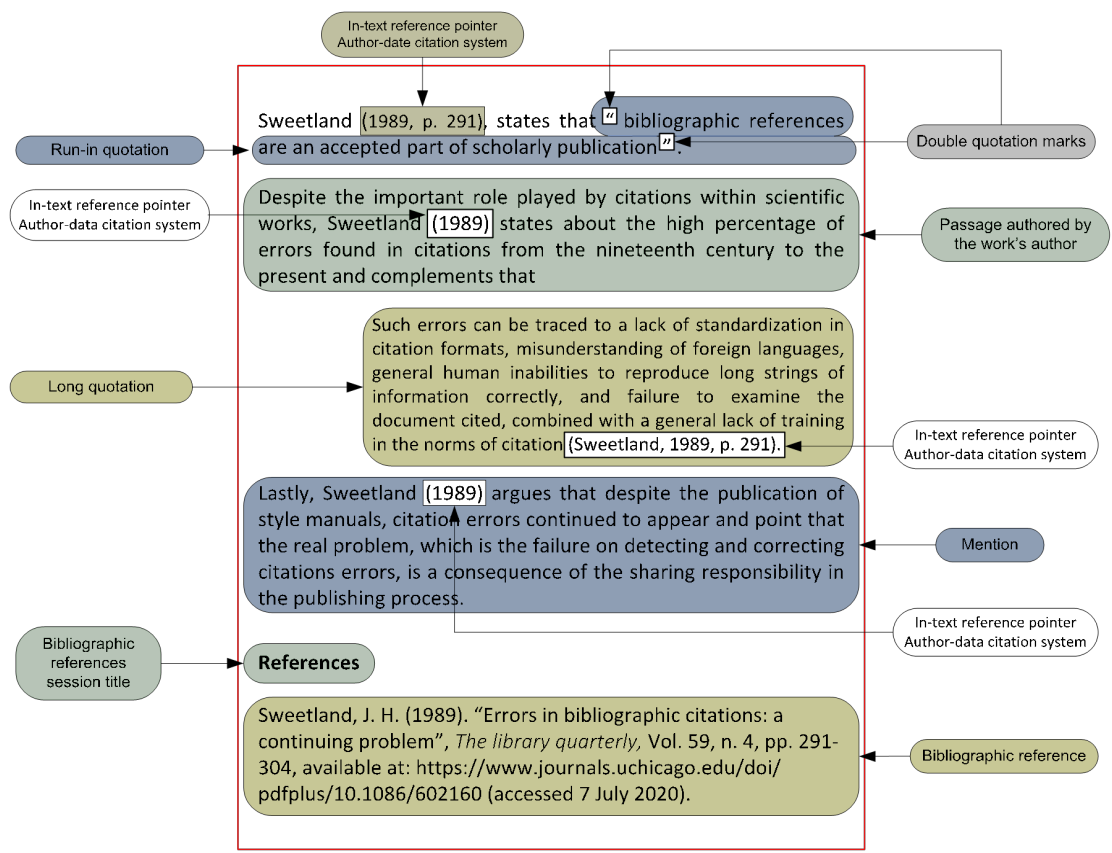

In particular, we aim at answering the following research questions (RQ1-RQ3 from now on):

RQ 1. Considering current bibliographic tools like reference styles and reference manager softwares, were they effective on fully addressing the issues pointed out by the study made by Sweetland in 1989?

RQ 2. Are there other possible causes for errors in citing and referencing other than those specified by Sweetland?

RQ 3. What impacts should be expected by readers when retrieving information from citing and referencing metadata, considering the current descriptive representation revision and the potential differences between the level of description adopted by bibliographic catalogs and bibliographic references?

**Figure 2.** Visual representation of the main bibliographic elements involved in citing and referencing citation-sequence system – i.e. when in-text reference pointers referring to mentions and quotations are specified using a number corresponding to a particular bibliographic reference in a bibliographic references list arranged in ascending numerical order

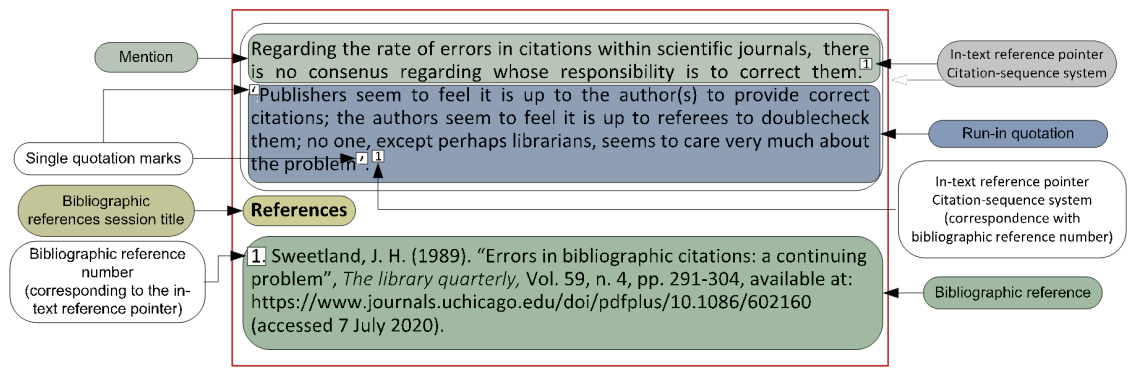

## Material and methods

We used the SCImago Journal & Country Rank (https://www.scimagojr.com/) to select a population of journals and their related articles for this study. SCImago enabled us to filter the journals it describes according to their subject headings which are based on those available in Elsevier's Scopus (SCImago, c2020; Guerrero-Bote and Moya-Anegón, 2012). What follows are the core points of the methodological procedures for the data collecting and analysis. More information about the methodology are introduced in Santos *et al*. (2020c).

As of 1 November 2019, SCImago contained 24.702 records of journals. Since we were interested in gathering information related to two subject areas, i.e. Medicine and Social Science, we looked only at the 7,224 records related to Medicine journals (17,93% of SCImago journals records) and the 5,855 records related to Social Sciences journals (14,53% of SCImago journals records). We used three criteria to select the journal titles to include in our sample: the "total cites" metric, the publisher, and the subject categories.

The first criterion, i.e. "total cites" metric in SCImago, corresponds to the number of citations received by the articles of a journal that have been published in 2015-2017, coming from the works (of any discipline) published in 2018. We considered this metric as a proxy for the prestige of journals – as pointed by Bonacich (1987, p. 1172), "high prestige is acquired by receiving unreciprocated choices from others". We ordered the journals in SCImago in decreasing order according to their "total cites" values to consider journals that are the most relevant ones in their particular subject area.

The publisher of the journal (i.e. the second criterion) was considered to have heterogeneous coverage. In particular, we avoided having more than one journal per publisher for each subject area to prevent scenarios where a publisher applies the same editorial policy for all its journals.

Lastly, we considered journal subject categories (i.e. the third criterion) for the selection. Each SCImago subject area is further classified in categories that define specific subheadings for a particular journal, e.g. Hematology for Medicine and Demography for Social Sciences. We selected only the journals that were classified, in SCImago, under a single subject area, and a single subject category to avoid possible overlaps between the selected journals in terms of areas and categories.

We decided a threshold date, i.e. 31 October 2019, as an upper bound to select journal regular issues (special issues were not considered) within the selected journals. In case a journal did not publish in issues, we considered, as a virtual regular issue, all the articles published in the previous month, i.e. between 1 October 2019 and 31 October 2019. For each journal issue selected which was not a special issue, we randomly chose 5 articles. Only the articles introducing original research communications were considered. If the selected issue did not contain all the 5 articles, we considered all the published articles within it.

Each article was individually analyzed considering quantitative and qualitative aspects concerning its mentions, quotations, and bibliographic references. We assumed that the citation behavior (i.e. the way mentions, quotations, bibliographic references, and in-text reference pointers are specified) is similar among different regular issues of the same journal.

## Data

Considering the data available on SCImago in November 2019, we selected 46 journals and 213 articles from both the subject areas in consideration, i.e. Medicine and Social Sciences. The articles contained a total amount of 9,911 bibliographic references and 16,193 mentions and quotations overall. We obtained 27 journals, 132 articles, 5,340 bibliographic references, 8,400 mentions and 5 quotations in Medicine, and 19 journals, 81 articles, 4,571 bibliographic references, 6,953 mentions, and 835 quotations in Social Sciences.

The sample size proportionally corresponds to the representativeness of both disciplines, i.e. Medicine and Social Sciences, within SCImago database. The method adopted for the selection of the journals allowed us to come up with a balanced and representative population of journals which supported us in doing reasonable initial refletions on citing and referencing issues, that may be extended in forthcoming studies.

The sample comprises journals owned by publishers from 13 different countries. Publishers of Medical journals are from Canada, China, Egypt, Germany, Hong Kong, Netherlands, New Zealand, Switzerland (each country with 1 journal representing 3.7% of the Medicine sample), United Kingdom (8 journals representing 29.6% of the Medicine sample) and United States (11 journals representing 40.7% of the Medicine sample). Publishers of Social Sciences journals are from Brazil, Germany, Portugal, South Korea (each country with 1 journal representing 5.9% of the Social Sciences sample), United Kingdom (7 journals representing 41.1% of the Social Sciences sample), and United States (6 journals representing 35.2% of the Social Sciences sample).

Gathered data included information to support a discussion from multiple points of view, i.e. the publisher, journal, and article perspectives. Also, we collected more granular data about in-text reference pointers and bibliographic references, which provided a view on the citation apparatus. These different viewpoints supported an overview of the variations with which bibliographic data appears and how they relate to each other in articles considering the subject areas we analyzed.

All the data used in our study are available in Santos *et al.*(2020b).

# Results

33% (i.e 9) of the Medicine journals and 35% (i.e. 6) of Social Sciences journals composing the sample are Gold/Diamond Open Access journals[2]. From an article perspective, 42% of the Medicine articles (55)[3] and 52% of Social Sciences articles (42) can be freely accessed from the journal website without paying any fee – because either the journal is a Gold/Diamond Open Access one or it is a Hybrid journal.

Analyzing the works cited by the articles included in our sample, we observed that 55% of the works cited by the Medicine articles (2915) are freely available online (either as Green, Gold or Diamond Open Access items) and 50.1% of the bibliographic references referring to them (1462) do not provide a DOI URL (e.g. an URL starting with either "http(s)://doi.org/" or "http(s)://dx.doi.org/" followed by a DOI) to access them directly from the Web. Along the same line, only 35% of the works cited by Social Sciences articles (1580) are freely available online and 61% of the bibliographic references referring to them (972) do not provide a DOI URL.

Considering all the bibliographic references included in Medicine articles providing a URL that did not include any DOI (even when hidden behind a hyperlink), 20% of them (818) specified a URL, while only 12% of bibliographic references in Social Sciences articles (324) contained an URL. However, 42% of the URLs indicated in bibliographic references of Medicine articles (342) often referred to records within a bibliographic database (e.g., a library's bibliographic catalog, like Pubmed[4]). We observed similar behavior in 2% of the bibliographic references (89) in Social Sciences articles.

The data we gathered included the reference styles adopted by the journals, according to the recommendations in the instructions provided for authors. We noticed a huge variety of reference styles among the journals in the sample, as shown in Table I.

70% of Medicine journals (19) adopted widely accepted reference styles like Vancouver, Chicago, AMA, and APA. Among these, 26.3% (5) customized the adopted reference style guidelines according to their specific needs. Such customization practice was not observed in any Social Sciences journal in our sample, which usually directly reused the original reference style chosen as it is.

Considering only journals providing own reference styles (Medicine: 26% - 7 journals; Social Sciences: 17.6% -3 journals), the reference styles of 33% of Medicine journals (8) and 23.5% of Social Sciences journals (4) did not provide clear, comprehensive, and exhaustive guidelines for accurately describing and arranging bibliographic metadata in bibliographic references. The reference styles can be classified into three non-disjoint categories:

1. those which did not provide guidelines for describing some types of publications like grey literature, e-prints, technical reports, speeches, etc.;

---

[2] See (Piwowar *et al.*, 2018) for a definition of all the Open Access levels.
[3] From now on, numbers between round brackets indicates the raw values represented by percentual indexes.
[4] PubMed is a free resource supporting the search and retrieval of biomedical and life sciences literature with the aim of improving health–both globally and personally (available in https://pubmed.ncbi.nlm.nih.gov/about/).

2. those which did not provide instructions on how to properly establish correspondences between bibliographic references and the in-text reference pointers used in mentions and quotations and;
3. those which did not provide instructions on how to proceed with particular bibliographic issues, like citing secondary or indirect sources (e.g. quoting quotations), structuring and formatting DOI metadata, etc.

**Graphic I.** Percentage of the adoption of reference styles in the journals in our sample

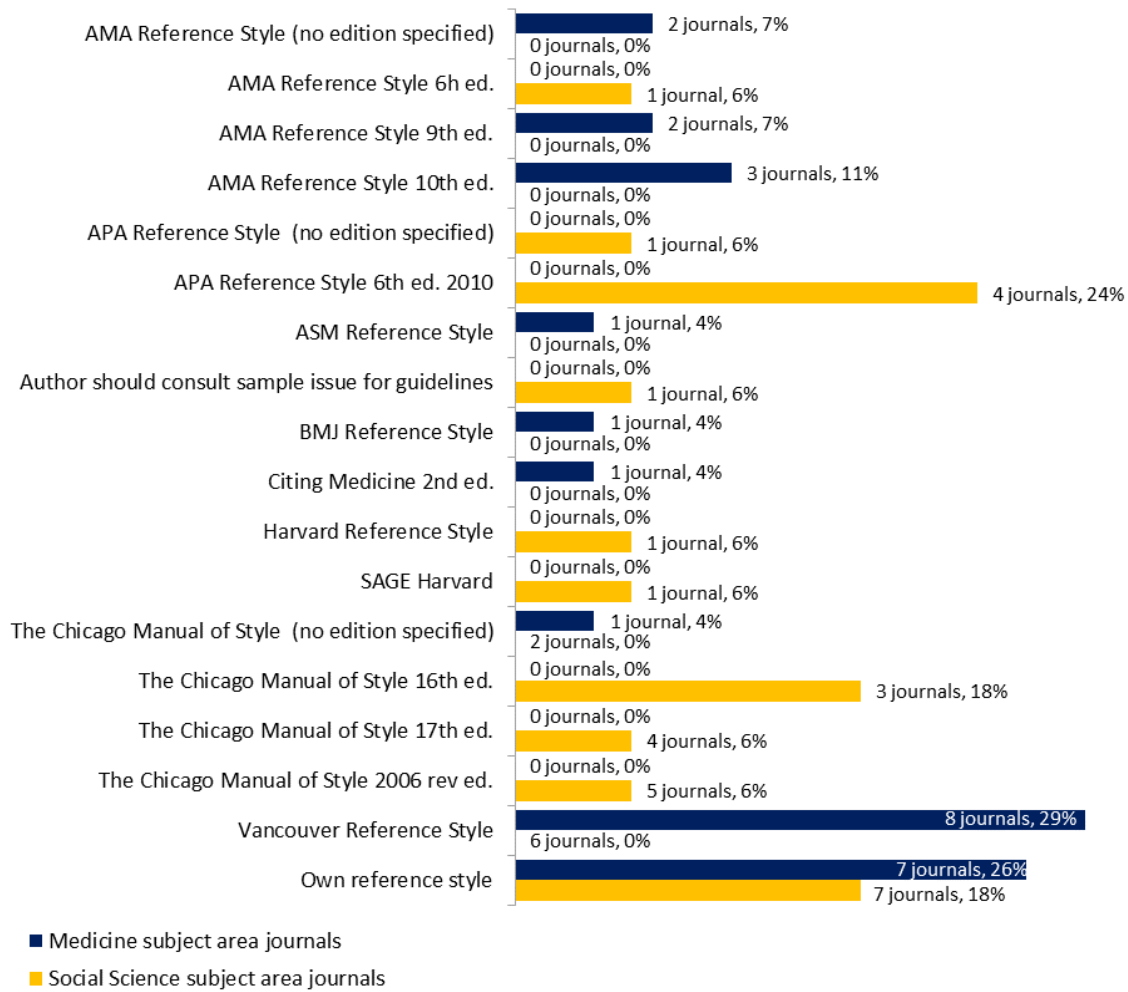

In Social Sciences articles, we noticed variations in the title attributed to the section containing bibliographic references. 87.6% of articles (71) name it *References*, 6.2% (5) name it *Referências* (References, in Portuguese, following the article's language), and 6.2% (5) name it *Notes*. 100% of Medicine articles (132) name the bibliographic references section as *References*.

The analysis of the type of publications cited by the articles in our sample showed 33 different types of works, as shown in Graphic II.

**Graphic II.** Percentage of the types of the cited works derived from analyzing the related bibliographic references (BRs) of all the articles in our sample

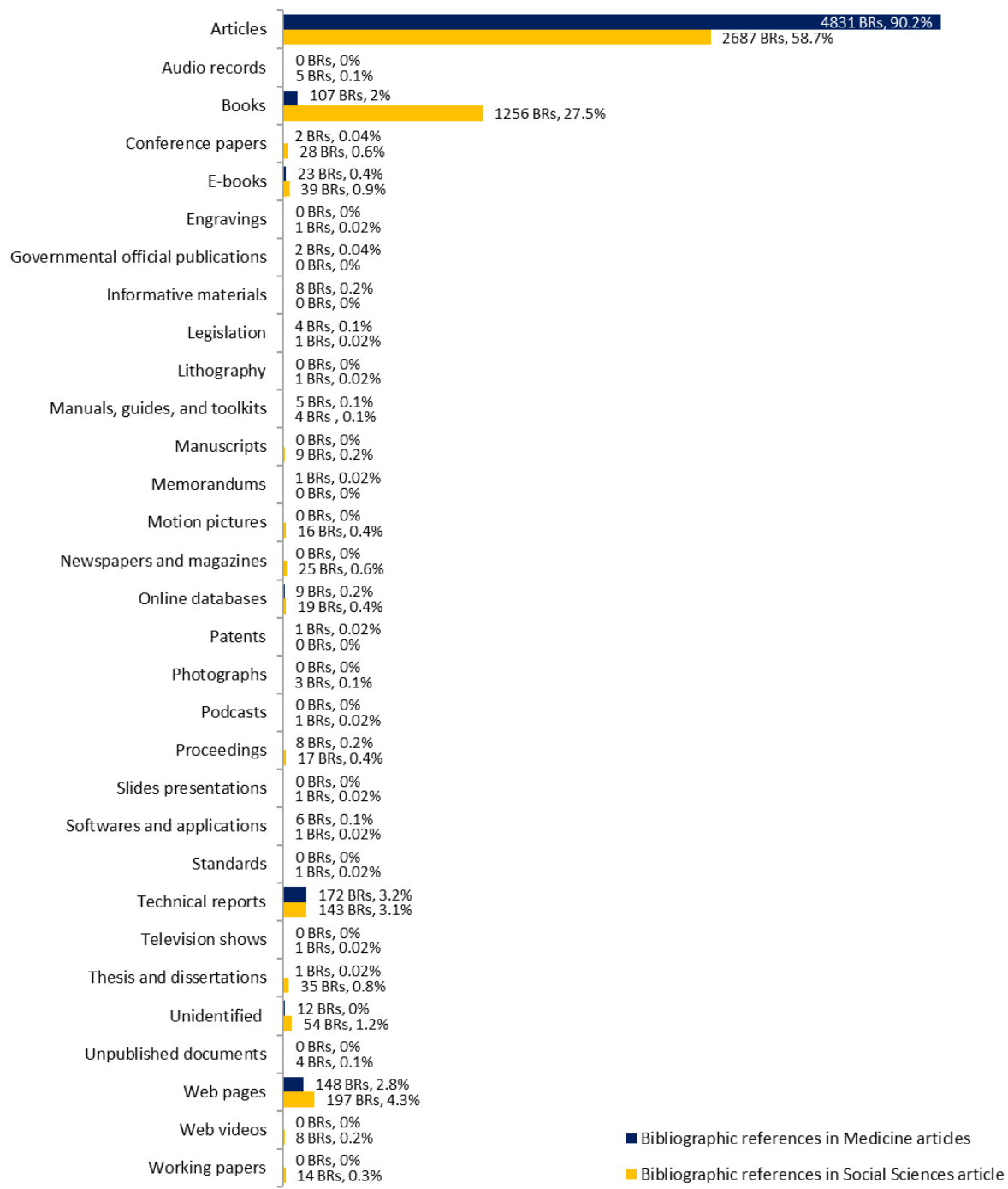

41% of Medicine journals (11) and 29.4% of Social Science journals (5) recommended the use of one or more reference management softwares. From this, 90% of journals in Medicine (10) recommended the use of Endnote (https://endnote.com/), 27% (3) recommended Mendeley (https://www.mendeley.com/), and 18% (2) recommended Zotero (https://www.zotero.org/). Considering Social Sciences journals, 100% (5) suggested Endnote, and 20% (1) suggested Zotero.

Some journals (Medicine: 89% (24); Social Sciences: 88.2% (15)) enabled one to download bibliographic metadata of their articles in textual or machine-readable formats, as shown in Table III.

From this, a portion of publishers (represented by 12,5% of Medicine articles (10)) provided bibliographic metadata for their articles in text format, with no information regarding the reference style in which it is formatted.

**Graphic III.** Percentage of the export formats of bibliographic records of the articles published in the journals in our sample. The "Text" category refers to journals providing bibliographic references in plain text with no specification of the reference style in which it is formatted).

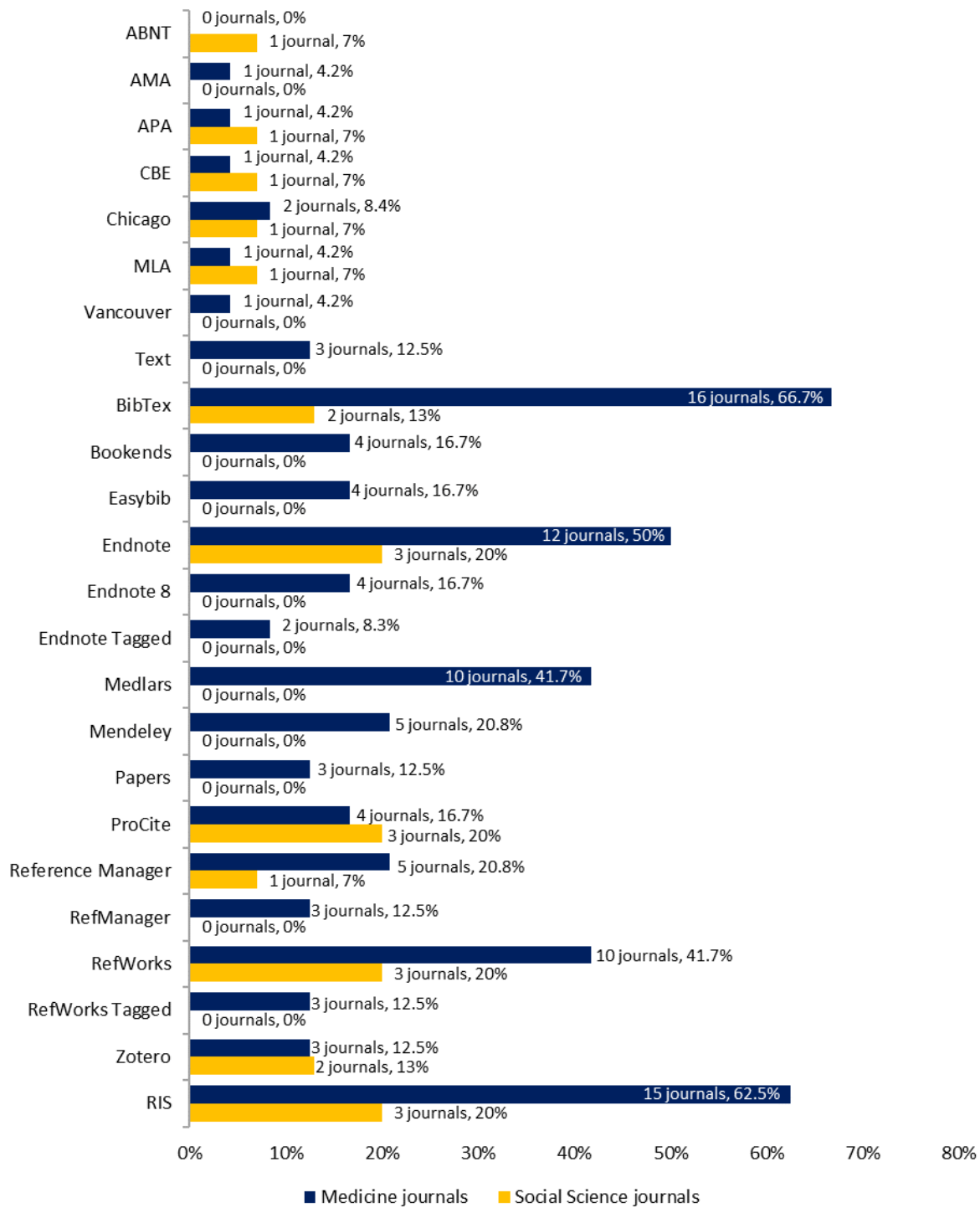

96% (136) of Medicine articles and 97.5% (79) of Social Science articles provided their own bibliographic metadata in headers, footnotes, or the first page of PDF files they provide.

Usually, articles adopt one out of two systems for identifying in-text reference pointers. The first one is the *author-date* system – e.g. "(Doe, 2020)" – adopted by 3.7% (5) of Medicine articles and 88.2% (71) of Social Sciences articles. The second, the *citation-sequence system*, based on numbers (e.g. "[3]"), was adopted by 96% (127) Medicine articles and 5.9% (5) of Social Sciences articles. The remaining 5.9% (5) of Social Science articles adopted both styles within the same articles. All these data are introduced in Table IV.

Considering articles adopting the citation-sequence system, we observed variations on the format of the numbers associated with each bibliographic reference, as shown in Table V. Within this system, the numerical arrangement of bibliographic references did not correspond to the order in which the respective in-text reference pointers appear in the text in 8.66% (11) of Medicine articles. For instance, let us consider the first paragraph of a Medicine article, shown in Figure 3. The first in-text reference pointer (e.g. "[41]") refers to the bibliographic reference in position 41 and the second one (e.g. "[40]") to the bibliographic reference in position 40. Similarly, the three in-text reference pointers that follow do not follow an ascendant numerical sequence: [1, 21, 31], [13, 27], [31, 34]. The journal in which this article was published adopts Vancouver reference style, according to which "references should be numbered consecutively in the order in which they are first mentioned in the text" (ICMJE, c2020), something that is not happening in this case. No additional instructions regarding the numerical arrangement, neither for in-text reference pointers nor for the bibliographic references, were found within the reference style.

**Graphic IV.** The format used for representing in-text reference pointers referring to mentions and quotations within articles

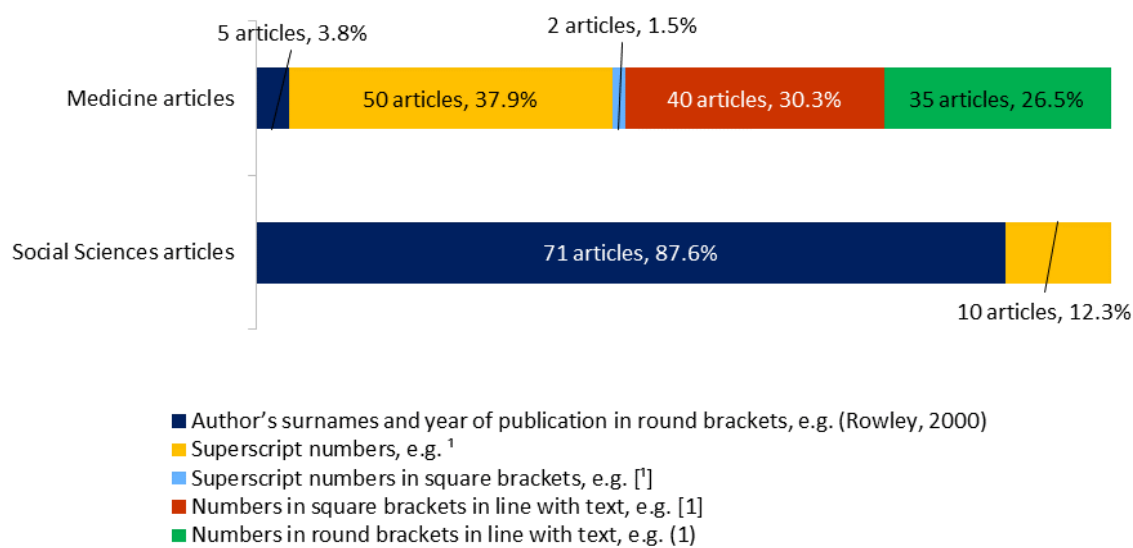

**Graphic V.** Different uses of the numbering systems for ordering bibliographic references in the bibliographic references list, considering articles adopting citation-sequence system.

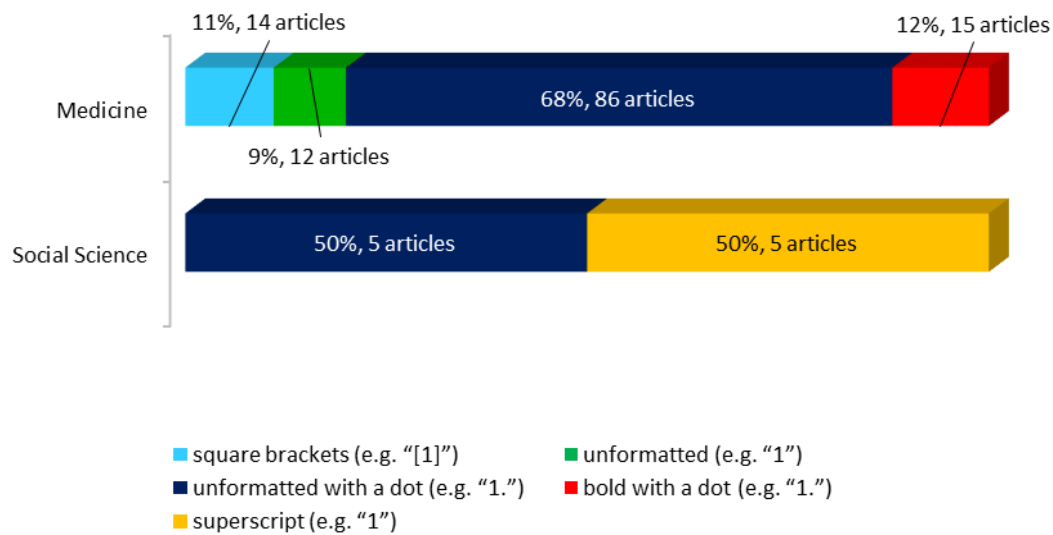

**Figure 3.** An excerpt from a paragraph in a Medicine article in our sample[5]

> The World Health Organization (WHO) defines mental health as "a state of well-being in which the individual realizes his or her own abilities, can cope with the normal stresses of life, can work productively and fruitfully, and is able to make a contribution to his or her community." [41] One in four people worldwide are affected by mental health concerns [40].

We observed misuses regarding the alphabetical arrangement of bibliographic references in 40% of Medicine articles (2) adopting such system and 3% of Social Sciences articles (2) adopting the author-date citation system. Figure 4 includes an excerpt of the bibliographic reference list of one of the articles in our sample, in which its bibliographic references were expected to be sorted in ascending order by authors' surnames.

Some publishers also provided hyperlinks to connect in-text reference pointers to the corresponding bibliographic references. This behavior was observed in 60% of Medicine articles (79) and 25% of Social Science articles (20). However, the reciprocal hyperlink, i.e. between a bibliographic reference to the in-text reference pointers denoting it, was not a usual feature provided in the Medicine articles. Instead, considering only the articles with hyperlinked in-text reference pointers, 100% of Social Sciences articles (20) provide backlinks to the in-text reference pointers, while only 10% of Medicine articles (8) implement such functionality.

Regarding the uniformity of descriptive metadata, journal titles of cited articles may appear in distinct forms within bibliographic references. These differences may exist either when comparing bibliographic references list of articles from different issues of the same journal or even between bibliographic references included in the same bibliographic references list. Differences regarded to the abbreviation of journal titles were observed only in Medical journals since 100% of Social Sciences articles (81) gave full titles of cited journals in their bibliographic references. Considering the 93% of Medicine articles (124) providing titles of cited journals in the abridged format in their bibliographic references lists, 4% (5) adopts the ISO 4 rules[6], 8% (10) adopts the ISSN List of Title Word Abbreviations (LTWA) guidelines[7], 1.6% (2) adopts the recommendations of the National Center for Biotechnology Information Database, 44.3% (55) adopts the recommendations of US National Library of Medicine (NLM, also referred by publishers as Pubmed Database and Index Medicus)[8]. For the remaining 42% (52) of Medicine journals, the source in which abbreviations should be based on was not identified.

---

[5]Passages referred in Figures 3 to 8 were extracted from articles in our sample.

**Figure 4.** An excerpt of the bibliographic reference list of an article in our sample

> Maparyan, L. (2012). The womanist idea. New York, NY: Routledge.
>
> Morris, E. W. (2007). ""Ladies" or "loudies"? Perceptions and experiences of Black girls in classrooms." Youth & Society, 55(4), 490-515.
>
> Ladson-Billings, G. & Tate, W.F. (Eds.). (2016) Covenant keeper: Derrick Bell's enduring education legacy. New York: Peter Lang.
>
> Johnson, L. (2017). The Racial Hauntings of One Black Male Professor and the Disturbance of the Self(ves): Self-Actualization and Racial Storytelling as Pedagogical Practices. Journal of Literacy Research, 49(4), 476-502.
>
> Lorde, A. (1984). Sister Outsider: Essays and Speeches. Trumansburg, NY: Crossing Press.
>
> Matthew, P.A. (2016). Written/Unwritten: Diversity and the hidden truths of tenure. Chapel Hill, NC: The University of North Carolina Press.

Our data revealed that articles did not usually provide sources of non-textual content – i.e. all that information presented by using visual signs, tables, graphs, photographs and images, illustrations, schemes, verbal communications, audio, and video recordings or any other type of manifestation made without using argumentative text as the main language. Non-textual content was observed in 92.4% of Medicine articles (122) and 71.6% of Social Sciences articles (58). Of these, 84.8% of Medicine articles (112) did not provide the source of non-textual content and 4% of articles (5) provided the source for only part of non-textual content. In the Social Sciences articles in our sample, 31% of them (18) provided the source of non-textual content and 12% of articles (7) provided the source for part of it.

---

[6] Refers to a Standard published by the International Standardization Organization (ISO), entitled "ISO 4:1997 Information and documentation — Rules for the abbreviation of title words and titles of publications" (https://www.iso.org/obp/ui/#iso:std:iso:4:ed-3:v1:en)

[7] Refers to a list containing standardized abbreviations used for words in scientific citations. (Available at https://www.issn.org/services/online-services/access-to-the-ltwa/?lang=en)

[8] Available at https://www.ncbi.nlm.nih.gov/nlmcatalog/journals

Quoting and mentioning proved to be more frequent in Social Sciences articles than in Medicine articles. In Medicine, we observed an average of 64 mentions and 0.04 quotations per article, while in Social Sciences we had 86 mentions and 10.31 quotations per article. 71.6% of Social Science articles (58) included at least one quotation, which seemed to confirm the impression that Social Sciences usually quote more compared with Medicine articles, in which only 2.27% of articles (3) used quotations.

About the markup used to show and identify quotations within a text, 100% of quotations in Medicine articles (5) used double quotation marks (i.e. "") to markup the text in run-in quotations. No long quotations were detected in Medicine articles in our sample. In Social Sciences articles, quotations were marked up in different ways. For run-in quotations, 70.6% of the articles (41) adopt double quotation marks (i.e. ""), 24.1% of the articles (14) adopt single quotation marks (i.e. ''), and 1.7% of the articles (1) adopt both double and single quotation marks. Considering long quotations, 31% of Social Science articles (18) used simple indentation, while 3.4% of the articles (2) presented quotations in indented passages and italicized characters between double quotation marks.

Usually, providing the page numbers in the in-text reference pointer associated with a mention is optional. However, it was used in 41.9% of the articles in Social Science (34), while Medicine did not show such behavior. Instead, the page numbers in in-text reference pointers referring to quotations were found in 68.9% of Medicine articles (2) and 68.9% of Social Science articles (40).

From a FRBR perspective, in-text reference pointers with page numbers were used to refer to a particular FRBR Manifestation of a cited document (i.e. a particular edition). Instead, the remaining in-text reference pointers, i.e. those without the specification of page numbers, actually pointed to the FRBR Expression level of the cited documents (i.e. their content), despite the particular format specified in the metadata of the related bibliographic references, which usually described the cited document at the Manifestation level.

The article analysis revealed that considering articles with quotations, 33.3% of Medicine articles (1) and 48.2% of Social Sciences articles (28) specified either in-text reference pointers or bibliographic references (or both) related to quotations that did not provide easy access to the text quoted. With *easy access,* we mean to locate a quoted passage within the cited document without having to perform complementary searches, like queries on indexes and summaries, or to read long excerpts to identify the quotation. One example of these cases is illustrated in Figure 5, which reproduces a passage from a Social Science article containing a quotation of a Collin's work, published in 2000. We noticed that the in-text reference pointer matches two different bibliographic references in the bibliographic references list, both published in 2000. Since the in-text reference pointer do not provide any specification of which bibliographic references it referred to, e.g. by adding an alphabetical character to the date of publication, like 2000a, and assuming that the provided pagination is correct, the only way of retrieving the quoted passage on the cited work is by consulting both works referenced by the two bibliographic references.

The excerpt in Figure 6, extracted from an article from the Social Science subject area, illustrates some of the issues mentioned above. In the original source, the year of publication of each cited work within the in-text pointer is connected to the respective bibliographic reference in the bibliographic reference list through a hypertextual link. By clicking on the link provided in the in-text reference pointer for "Okoh and Hilson, 2011", we were sent to the bibliographic reference shown in Figure 7.

**Figure 5.** An excerpt from a Social Sciences article

> In the text-body:
>
> It also develops space for Black women researchers to do work that Collins (2000) describes as activating "epistemologies that criticize prevailing knowledge and that enable us to define our own realities on our own terms..." (p. 292).
>
> In the bibliographic references list:
>
> Collins, P.H. (2000). Black feminist thought: Knowledge, consciousness, and the politics of empowerment. New York, NY: Routledge.
>
> Collins, P. H. (2000) "What's going on? Black feminist thought and the politics of postmodernism." In E. A. St. Pierre & W.S. Pillow (Eds.) Working the ruins: Feminist post-structural theory and methods in education (41-73). New York: Routledge.

**Figure 6.** An excerpt of a Social Sciences article

> "This is the case for many countries such as Ghana, Tanzania, Senegal, and Mozambique (Fisher, Mwaipopo, Mutagwaba, Nyange, & Yaron, 2009; Aizawa, 2016; Bryceson & Geenen, 2016; Hilson & Garforth, 2012; Okoh and Hilson, 2011; Persaud, Telmer, Costa, & Moore, 2017)."

**Figure 7.** The bibliographic reference of one of the in-text reference pointer introduced in Figure 6

> Hilson, G. (2011). Artisanal mining, smallholder farming and livelihood diversification in rural Sub-Saharan Africa: An introduction. Journal of International Development, 23(8), 1031–1041.

None of the works authored by Hilson described in the bibliographic reference list was co-authored with Okoh. The only work dated from 2011 considered in the bibliographic references list was the one reproduced in Figure 7. By checking the work represented in the bibliographic reference in the webpage of its publisher, we confirmed that Okoh was not an author of the work referred by the bibliographic reference. Thus, there is no evidence that the in-text reference pointer and bibliographic reference referred to the same work, neither that the quoted passage is actually a quotation or a mention incorrectly marked up as a quotation.

Figure 8 illustrates a situation in which the author does not use the same approach to specify in-text reference pointers within the text. Indeed, since the in-text reference pointer referring to the bibliographic reference number 41 appears after the period ending a quotation and no page number was provided in it, readers have to infer autonomously to which passage in-text reference pointers 40 and 41 refer to since it is ambiguous whether 41 refers to the previous sentence or not. These are only a few examples demonstrating how ambiguous bibliographic metadata was within the articles in our sample.

**Figure 8.** Another excerpt from a Social Science article.

> …community". [41] One in four people worldwide are affected by mental health concerns [40].

## Discussion

In this section, we discuss the results introduced in the previous sections focusing on answering the four research questions (R1-R4) presented in the introduction. In particular, the discussion is organized in two parts. In the first part, we address RQ1 and RQ2, while in the second part we discuss RQ3.

### Issues in citing and possible causes for errors

90.23% of Medicine bibliographic references refer to articles published in journals. This is probably due to the workflow of journals which is relatively more dynamic compared to books, which favors fast discussions in health sciences. Instead, in Social Sciences domains, aspects of sociological, historical, cultural, political, chronological, anthropological, and geographical nature, directly influencing research trends are approached both in articles and in other types of publications, such as books. This justifies the greater variety of typology of works cited by social sciences articles in comparison to medicine articles according to data shown in Table 2.

While usually published in restricted access journals, Medicine articles of our sample showed that they cited more freely available articles than restricted access ones, even when such cited articles were published in restricted access journals. However, from a Social Sciences perspective, we noticed the opposite behavior, since Social Sciences articles in our sample tended to cite restricted access articles without providing any metadata for their freely available version (if any).

Providing URLs and hyperlinks within bibliographic references was a frequent behavior in Medicine articles. However, the accuracy of the location pointed by such URLs/hyperlinks was more reliable in Social Science articles. Indeed, the number of times an URL/hyperlink pointed to a wrong Web location, e.g. to an online bibliographic catalog instead of the cited work itself, was higher in Medicine articles. This suggests two things. First, Social Science author's awareness of promoting the access to freely available content may result in the higher reliability of the hypertext links provided in their bibliographic references. Second, we did not find detailed instructions regarding the provision of URLs and DOI hyperlinks within bibliographic references in reference styles (RQ1). In particular, while usually encouraging authors to provide these bibliographic data, the reference styles analyzed generally did not indicate clear guidelines regarding the description procedures neither on the location where those hyperlinks should point to, such as the actual file containing the cited work nor to the landing page from where it may be downloaded (RQ2).

These behaviors suggest a resistance – by authors, editors, or both – on providing enough metadata in the bibliographic reference to facilitate the access to the cited works (RQ2). Although the articles' content and its accuracy are authors' responsibilities, there did not seem to be an adequate verification of the bibliographic references before their publication, which may suggest that the efforts devoted by publishers of the documents in our sample in providing trustful bibliographic data in their articles, especially concerning bibliographic references, were not as meaningful as those related to content quality. In this sense, Sweetland (1989, p. 300) considers that "the role of citations is not taken very seriously by the scientific community". Cronin (1982, p. 71) complements that "journal editors and referees could pay greater attention to the quality and quantity of references".

Our analysis showed the adoption of 20 different reference styles within the journals of the sample. According to Table 1, the average number of reference styles adopted by journals composing the sample is 45% higher in Social Sciences journals than what we observed in Medicine journals. Such diversity in the adoption of reference styles in the same subject area weakens the argument that the existence of multiple reference styles is justified by the specific needs in different disciplines (Gratz, 2016), specific rules for a particular audience (Barbeau, 2018) and by tradition (Barbeau, 2018; BibMe, 2017; Gratz, 2016), for instance:

> "social sciences tend to use current research, so the publication date of a source is very important. For this reason, people in the social sciences tend to use APA style; APA style puts the date before all other information, aside from the author's name, which makes it much easier for researchers in that field to find valid, up-to-date information." (UCM Writing Center, 2016)

This huge variety of reference styles disfavors the uniform bibliographic metadata description in bibliographic references since different reference styles can recommend different ways of representing the same information. As stated by Sweetland, "complaints about lack of uniformity are common in the literature, whether from authors or librarians". The author complements that "given the variety of formats for citation and the lack of any real agreement among journals or authors, the chance of misunderstanding is high" (Sweetland, 1989, p. 298).

Some journals recommend more than one bibliographic reference managers within the instructions provided to authors for writing their manuscript. Updating the format used by such reference managers to return bibliographic references according to the reference styles' guidelines can prevent journals from receiving differently formatted articles. Theoretically speaking, that could decrease the editorial work on checking bibliographic metadata accuracy.

From the reference manager administrator's perspective, the more the number of reference styles available is huge, the greater the challenge of creating algorithms for recognizing and stylesheets for formatting bibliographic references appropriately. Besides, an incredible effort is needed to update such algorithms and stylesheets when, for instance, particular reference styles have variations on their citing systems (Barbeau, 2018) and multiple versions and editions of the same reference style are introduced in time.

The diversification of guidelines may confuse researchers. A clear example of this situation is the name used by publishers to refer to reference styles in their homepages. According to the data we gathered, we noticed several variations – e.g. AMA Reference Style was often mentioned as AMA Manual of Style, Harvard Reference Style was also referred to as Harvard Reference System, and Vancouver Reference Style was also referred as ICMJE and Citing Medicine.

We also noticed that the name attributed to the section containing bibliographic references in Social Sciences articles may change. For example, sections containing bibliographic references are named differently within articles adopting the 16th edition of Chicago reference style: 93.8% (76) name it "References" and 6.2% name it "Notes" (5). Besides, two different citation styles were observed on this same sample slice: citation-sequence adopted by 87.6% of articles (71) and author-date adopted by 6.2% of articles (5). The remaining portion of articles adopts both citation systems. No misuses of adopted reference styles were observed in medicine articles, considering specifically these aspects.

We speculate that, in time, failures regarding the interpretation of bibliographic guidelines description cannot be exclusively attributed to the availability or use of multiple reference styles (RQ2). Indeed, our analysis revealed that reference styles content could be clearer, including the ones authored or adapted by journal publishers. We observed shortcomings and omissions, as discussed in the results session, which may increase the probability of making mistakes like those introduced in Figures 3-8, as well as those pointed by Sweetland (1989).

Most of the Medicine journal publishers in our samples adapted reference style guidelines to their particular purposes. However, even with these changes, the reference styles remained vague and imprecise in some respects and may result in errors in mentioning, quoting, and referencing cited works (RQ2).

Similarly, although Social Sciences publishers generally did not recommend any adaptation to the existing reference styles adopted by their journals, often the guidelines for writing and formatting bibliographic references and in-text reference pointers related to mentions and quotations were not clear and easy to understand. For instance, while the most adopted reference styles focused on instructions for bibliographic references content and formatting, they generally did not provide guidelines for describing all the types of publications cited in the articles in our sample such as engravings and lithography (RQ1). These shortcomings, also mentioned by Sweetland (1989), often are the reason that denies a precise identification of the cited works by a reader. This situation seemed particularly relevant in Social Sciences articles, which cited many types of publications.

Also, 5.9% of the Social Sciences journals (1) did not adopt explicitly a reference style and, instead, recommended the authors to consult the bibliographic references of the journal sample issue and to consider them as a model for writing and formatting bibliographic references. However, the coverage of the types of cited works in such sample issue was not complete. In these cases, the author had no alternative unless writing the bibliographic reference according to his/her own beliefs on what could be the best way of describing the cited works. This may explain some cases in which required metadata for the identification of the cited work are actually missed (RQ2).

Along the same lines, we observed that some of the reference styles provided by the publishers of the Medicine articles in our sample, which adopted a citation-sequence system, did not mention the rationale in which the bibliographic references should be sorted in the bibliographic reference lists, thus creating even more confusion for the author who wrote them. Also, we observed that four variations of the style of the numerical character of such bibliographic references adopted in citation-sequence system in Medicine articles, while for Social Science articles there were found only two variations.

Regarding how quotations are marked up within the article content, we found cases within Social Science articles in which double and single quotation marks were adopted for run-in quotations, while long quotations were indicated using indentation which was, sometimes, accompanied by italicized characters and double quotation marks. Besides, we noticed that there were no shared rules used to identify a quotation as run-in or long. When specified, the main strategies adopted in the reference styles of the journal in our sample were:

1. to classify quotations considering the total numbers of words quoted (usually, 80 words at maximum for run-in quotations);
2. to classify quotations according to the length of the quoted passage in the citing work (usually, 3 or 4 lines at maximum for run-in quotations).

Considering the bibliographic references included in Medicine articles in our sample, the names of journals in which cited articles were published are provided in abridged formats, which are usually based on lists providing standardized abbreviations to journal names. In the meantime, we observed five different sources from where these abbreviations can be extracted. In principle, the same journal can be abridged in at least five different ways, which reinforce the claim by Sweetland (1989, p. 298) in which "differences in journal title abbreviation have been commonly noted as a particular source of error" (RQ2). Figure 9 reproduces one of the bibliographic references of an article composing our sample (Uzunalli *et al.,* 2019). In this example, the abbreviation "Ann" can be understood by the reader as "Annals", "Annual", "Annalen", "Annales", and so on. As an additional source of ambiguities, the guidelines for authors provided in the journal's webpage did not mention the source from which journal titles abbreviations should be based on.

**Figure 9.** A bibliographic reference with an abridged representation of the cited journal.

> Fanning AS, Anderson JM. Zonula Occludens-1 and -2 Are Cytosolic Scaffolds That Regulate the Assembly of Cellular Junctions. Ann N Y Acad Sci. 2009; 1165:113–20.
> https://doi.org/10.1111/j.1749-6632.2009.04440.x. [PubMed]

Thus, abridging journal titles within bibliographic references in Medicine articles seemed not to be fully accomplished. Thus, one possible strategy to adopt to disambiguate the journal would be to include complimentary information within bibliographic references, e.g. the cited journal's ISSN, to assure the uniform interpretations of journal titles abbreviations.

Our analysis also revealed a high percentage of articles not including the source of non-textual content included in the articles, e.g. tables and figures, especially in Medicine articles. Although non-textual content is also citable, we observed that reference styles rarely provide guidelines on how to proceed with the presentation and description of this kind of content.

Summarizing, all the information presented so far allowed us to conclude that the issues raised by Sweetland's' study in 1989 are still valid today (RQ1). We do not need thousands of reference styles if we do not have clear guidelines on how to inject bibliographic metadata into bibliographic references. In this context, neither the standardization nor the technologies developed were able to fully support the creation and management of bibliographic metadata to write bibliographic references. Reference styles do not provide sufficient and clear information to authors regarding the procedures of citation, metadata description, and formatting. Looking at the articles in our sample, reference styles seemed to be a list of suggestions used by authors and publishers to support their own decisions regarding referencing and citing rather than a concise and precise instrument of guidance compliant with standardized behaviors on citing habits.

As anticipated by Galvão (1998), our analysis confirmed that the terminology used to refer to bibliographical concepts was ambiguous. Terms like "reference styles" and "citation styles", and "bibliography", "bibliographic references", "references", and "notes" referred to the same thing on different occasions, thus confirming Sweetland's statements regarding the "lack of uniformity in the literature" and "of real agreements among journals and authors" (Sweetland, 1989, p. 298).

The rate of errors in citations in respected scientific journals in Medicine and Social Sciences is high (RQ2). Sweetland (1989) states that there is little consensus about who should be responsible for correcting citations: publishers may think it is up to the authors while authors would like to have referees and editors to double-check them.

In fact, there are no explicit rules or clear statements on the roles expected to be played by different agents in the editorial field (i.e. authors, editorial boards and publishers), nor behavioral patterns concerning citing and referencing matters among editorial market agents. For instance, some publishers provide an expert service for normalization and, in such cases, submissions are usually accepted in any formatting format. In other cases, publishers, that accept submissions in any format, provide authors with the adopted reference styles' guidelines just in case of acceptance of the submitted paper for publication. Again, in other situations, publishers indicate either a particular reference style or a set of guidelines and then leave the primary responsibility for normalization on the authors. In this situation, the limits concerning the purview of the of agents who are expected to directly act on normalization matters are tenuous and, therefore, not precise. This probably contributes to the confusion on the definition of the role to be played by the main core agents involved in the production of scientific content, as previously stated by Sweetland, who complements:

> "no one, except perhaps librarians, seems to care very much about the problem. [...] We spend considerable time and effort in training catalogers in both the theory and the methods of descriptive cataloging. It would be good to spend at least some effort on training all information workers [including authors and publishers] in the theory and methods of the citation." (Sweetland, 1989, pp. 301-302).

*Conceptual representation of references and citations*

Toward the end of the 20th Century, librarians began discussing possible changes in the description, access, and encoding of bibliographic information. Typically, especially in bibliographic catalogs which were based on the Anglo American Cataloging Rules (AACR), documents were described out of context and their descriptions usually referred to a particular edition published by a specific publisher. In modern times, this approach has been no longer sustainable and compliant with the fulfillment of the functions of bibliographic catalogs in the new information scenario (Joudrey *et al.*, 2015; Tillet, 2003b, OCLC, c2020).

Between 1992-1995, the IFLA Study Group on Functional Requirements for Bibliographic Records (FRBR) developed an entity-relationship based on a conceptual model for describing bibliographic records for all types of materials. This conceptual model should not only consider the function of the catalogs, which should enable users to find, identify, select, obtain bibliographic resources and navigate within the catalog, but also allow the performance of user tasks associated with bibliographic resources and the conceptual model of the bibliographic universe: the entities, their relationships, and attributes (Tillet, 2003a, 2003b; IFLA, 2009). FRBR determines that items (embodiment of works in physical or electronic publications) must be described in a context in a manner sufficient to relate the item to the other items comprising the work using the four-level bibliographic structure: FRBR Work, FRBR Expression, FRBR Manifestation and, FRBR Item (OCLC, c2020).

Although IFLA Functional Requirements family (FRBR, FRAD, FRSAD) was consolidated by the IFLA Library Reference Model (IFLA LRM) in 2017, we decided to consider the FRBR entities concepts in this study. FRBR concepts are needed to construct a theoretical background that supports further discussions concerning citing and referencing matters – indeed IFLA LRM concepts are more comprehensive than those of FRBR family. In addition, FRBR was the first concrete IFLA initiative that have changed descriptive representation approaches on information and, theoretically, may be considered the starting point of the distancing between the facets which compound descriptive representation.

Even adopting new perspectives, the essence of descriptive representation in bibliographic catalogs still focus on describing a publication as stored in a particular information support, be it physical or electronic. Indeed, despite all these new aspects introduced by FRBR, cataloging, and the preparation of bibliographic references, according to IFLA Study Group on the Functional Requirements for Bibliographic Records (2009), should aim at:

a) identifying all the background information that have supported an author's ideas conception, or the information which reading is being recommended by him as a complimentary content of his ideas;
b) using the information contained in the bibliographic reference (e.g. an author, the title, or a journal) to select an entity that is appropriate to one's needs;
c) finding the entities described.

Since the new approach of descriptive representation materialized in the FRBR guidelines considers document content (FRBR Expression) rather than the format in which it is embodied (i.e. FRBR Manifestation or, more concretely, FRBR Item), the relationship among mentions, quotations, bibliographic references, and in-text reference pointers seems to collide to some degree.

Quotations are excerpts of textual contents available in the cited works (FRBR Expressions), while mentions are the author's written interpretations of a textual content of the cited works (again FRBR Expressions) that should convey somehow the original idea of the cited author (FRBR Work). Instead, bibliographic references usually focus on one particular embodiment (FRBR Manifestation) which could be available in a single and physical (or electronic) exemplar of a publication (FRBR Item), e.g. the copy of a PDF file.

In this context, it should be considered that a FRBR Expression can be (and actually is) published in different information supports and formats. Since most of the scientific production is available on the Internet (sometimes free of charge) and since this content can assume different FRBR Manifestations, referring to specific FRBR Manifestations within bibliographic references may restrict the possibilities of access for the reader, if he does not have the perception (reading the bibliographic reference) that certain content to which he does not have access to (e.g. in the publisher website) may be available in a different FRBR Manifestation (e.g. in a preprint server). Besides that, in-text citations usually refer to the content of a bibliographic resource, that is, the FRBR Expression of a FRBR Work, while the respective bibliographic references usually refer to its embodiment (FRBR Manifestation). This richness represents also an obstacle to facilitating and improving the ways of accessing information, which also corresponds to one of the purposes of descriptive representation.

In our sample of Medicine and Social Sciences articles we found that the guidelines above were not always followed (RQ3). In particular, the descriptive elements provided in the main part of in-text reference pointers that accompany quotations pointed to the FRBR Manifestation layer of the cited work due to the presence of the pages where the quoted text is contained. However, a smaller portion of such in-text reference pointers did not provide any descriptive element for referring to the FRBR Manifestation layer and actually seemed to relate to the pure content of the cited works (i.e. their FRBR Expressions). Besides, we found that the main part of bibliographic references referred to a particular FRBR Manifestation of the cited works. However, there were a few cases where it was not possible to identify even the FRBR Expression of the work defined by a bibliographic reference. For instance, the bibliographic reference "World's Work. 1909. "The March of Events."". World's Work, December." found in one of the articles analyzed did not provide enough descriptive information to classify it properly.

Usually, the reference styles of the journals considered in our analysis encourage authors to provide page numbers in the in-text reference pointers referring to quotations, which can be used to locate the excerpt within the cited work if we strictly consider the particular FRBR Manifestation of the cited article referenced by the particular bibliographic reference. Although considering a broader context where each cited work can be fully characterized according to FRBR, that scenario can be restrictive, since does not allow one to consider the possible various kinds of print-like embodiments (printed within a volume, PDF in the publisher website, PDF in an institutional repository, etc.) the work may assume. In addition to that, we found there are no clear guidelines when the cited works cannot be paginated, such as in HTML versions of articles, speeches, digital media content, and tridimensional objects.

The evolution that has characterized the universe of information sciences may have brought some additional challenges in specific processes, such as the bibliographic normalization activity. In this work, bibliographic normalization was addressed as a facet of descriptive representation. Therefore, it has been inevitable to consider the evolutions and improvements that have been carried out in the cataloging domain, including the introduction of FRBR. Conceptual changes in cataloging should potentially impact the way bibliographic metadata is written and managed, the relations between in-text reference pointers, bibliographic references and the cited works, and bibliographic catalogs. Cataloging description level concept under FRBR, which corresponded to the descriptive comprehensiveness, needs to consider different aspects of the same work, which may switch the interpretation of the metadata available in the catalog. This is something to be considered in bibliographic normalization activities, to understand and, eventually, foresee how these changes can impact the way bibliographic references should be presented (RQ3).

## Conclusion

Our findings suggest that reference styles do not fully accomplish with their role of guiding authors and publishers on providing concise and well-structured bibliographic metadata within bibliographic references to allow the easy and accurate identification of the referenced works, in particular when different types of work, in addition to articles and books, are considered.

In particular, in response to RQ1 concerning the claims pointed by Sweetland (1989), our study confirmed that even with the use of bibliographic tools like reference styles and reference manager softwares, several of the issues highlighted by Sweetland (1989), introduced by table 1, still hold today. For instance, the variety of reference styles and the errors which are not corrected before publication support the speculation that journal editors did not seem to spend a huge effort on revising bibliographic metadata within their articles. We also detected possible causes for the lack of standardization on citing and referencing, other than those claimed by Sweetland (1989), to which RQ2 refers, such as the shortcomings, customizations and omissions within reference styles. In some cases, the importance of having clear and clean citation apparatus seemed not being a primary concern by both the authors and the publishers. Besides, the training of authors in bibliographic reference writing and in-text mentions and quotations of passages of a cited works seemed poor.

Since failures pointed by Sweetland (1989) were not overcome but incremented, as highlighted in our study, we suggested that the revision by which descriptive representation is going through should also be extended to comprise the citing and referencing normalization domains. In response to RQ3, which refers to the impacts that may be expected by readers on retrieving information from citing and referencing metadata considering such a revision, a trend towards an asymmetry on the way information may be represented within bibliographic references and bibliographic catalog was detected. In particular, the adoption of FRBR principles in bibliographic catalogs may affect the way bibliographic records are stored since each catalog could follow its own approaches to describe a certain document (e.g., only the FRBR Manifestation level of the first edition) which may differ from the others.

Finally, this study was based on the analysis of bibliographic metadata of a subset of articles of Medicine and Social Sciences subject areas. In the future, and on the basis of the results obtained in the current study, we will perform a broader and more accurate study in which we will complement the insights presented here by considering other subject areas, such as the Humanities.